\documentclass[12pt,a4paper]{article}
\usepackage[left=1in, right=1in, top=1.4in, bottom=1.3in]{geometry}
\usepackage[utf8]{inputenc}
\usepackage{amsmath}
\usepackage{amsfonts}
\usepackage{amssymb}
\usepackage[english]{babel}
\usepackage{graphicx}
\usepackage{url}
\usepackage{setspace}
\usepackage{natbib}
\bibpunct{(}{)}{,}{a}{}{;}

\newcommand{\blind}{1}

\newlength{\normalparindent}
\raggedright
\newcommand{\cov}{\hbox{Cov}}
\newcommand{\var}{\hbox{Var}}

\newcommand{\momtwo}{\mu_x^{(2)}}

\newcommand{\varyy}{\sigma_y^2}
\newcommand{\varx}{\sigma_x^2}
\newcommand{\varr}{\sigma_R^2}

\newtheorem{theorem}{Theorem}[section]

\begin{document}
\def\spacingset#1{\renewcommand{\baselinestretch}%
{#1}\small\normalsize} \spacingset{1}
\if1\blind
{
  \title{\bf  Sample Design for Medicaid and Healthcare Audits}
  \author{Michelle Norris \thanks{
   Michelle Norris is Associate Professor, Department of Mathematics and Statistics, California State University, Sacramento, 6000 J Street, Sacramento, CA 95819 (e-mail:  norris@csus.edu).The author thanks Martin Hauser for his careful proofreading of the manuscript and several helpful discussions.}\hspace{.2cm}\\
    Department of Mathematics and Statistics\\
     California State University, Sacramento\\
    }
  \maketitle
} \fi

\if0\blind
{
  \bigskip
  \bigskip
  \bigskip
  \begin{center}
    {\LARGE\bf Sample Design for Audit Populations}
\end{center}
  \medskip
} \fi

\bigskip

\abstract{We develop several tools for the determination of sample size and design for  Medicaid and healthcare audits. The goal of these audits is to examine a population of claims submitted by a healthcare provider for reimbursement by a third party payer to determine the total amount of money which is erroneously claimed. For large audit populations, conclusions about the total amount of reimbursement claimed erroneously are often based on sample data.  Often, sample size determination must be made in the absence of pilot study data and existing methods for doing so typically rely on restrictive assumptions. This includes the `all-or-nothing errors' assumption which assumes the error in a claim is either the entire claim amount or none of it.  Under the all-or-nothing errors assumption, \cite{roberts1978} has derived estimates of the variances needed for sample size calculations under simple expansion and ratio estimation. Some audit populations, however, will contain claims which are partially in error -- for example, a claim for a single patient visit to a physician may contain several line-item charges and only some of these may be in error. We broaden existing methodology to handle this scenario by proposing an error model which allows for  partial errors by modeling the line-item error mechanism. We use this model to derive estimates of the variances needed for sample size determination under simple expansion and ratio estimation in the presence of partial errors.  In the absence of certain error-rate parameter estimates needed to implement our method, we show that conservative sample sizes can be determined using the claim data alone.  We further show that, under all-or-nothing errors, ratio estimation will tend to outperform simple expansion and that optimal stratification is independent of the population error rate under ratio estimation.  The proposed sample design methods  are illustrated on three simulated audit populations.}

\vspace{1cm}
word count: 6490 words (LaTex code excluding tables and figures) 

\newpage
\spacingset{1.45} 

\section{Background and Motivation} \label{Intro}
According to the Medicaid program website (\cite{CentersforMed2018B}),

\begin{quote}
``Medicaid provides health coverage to millions of Americans, including eligible low-income adults, children, pregnant women, elderly adults and people with disabilities...The program is funded jointly by states and the federal government."
\end{quote}
In 2016, \$566 billion in Medicaid payments were disbursed to healthcare providers such as pharmacies, medical offices, and school districts in the US (\cite{CentersforMed2018A}). In California, MediCal is the name for the Medicaid program, and the California State Controller's Office is charged with conducting audits to ensure that MediCal funds paid to organizations conform to the requirements of the MediCal program and are of the appropriate amount.  

In planning a MediCal audit, auditors typically have access to a population of MediCal claims which they are charged with auditing for correctness. For example, if the audited organization is a medical clinic, a single claim may represent a single visit by a single patient, and the population may contain a million claims from a three-year period. The population may account for tens of millions of dollars in disbursed MediCal payments.  Because a complete examination of all claims is not feasible, auditors typically select a sample of claims, then, based on documentation, determine the appropriate amount of MediCal reimbursement that should have been paid for each claim in the sample. There are three possible outcomes for each sampled/audited claim: 
\begin{enumerate}
\item None of the amount claimed is disallowed, and the entire claimed amount is deemed allowable for reimbursement  (as shown in lines 1 and 5 in Table \ref{auditDataTable}).
\item The entire amount claimed is deemed disallowed, and none is deemed allowable for reimbursement (lines 3 and 6 in Table \ref{auditDataTable}).
\item  A portion of the total amount claimed is deemed disallowed and only the remaining portion is allowable for reimbursement(lines 2 and 4 in Table \ref{auditDataTable}). This  case is also called a partial payment or partial error. 
\end{enumerate}
A common assumption in the existing literature on audit sample design is the `all-or-nothing error assumption' which states that the error/disallowed amount in a claim equals the entire claim amount or zero.  The all-or-nothing error assumption precludes the possibility of partial errors but greatly simplifies theoretic calculations.\\

While the claim amounts are known for the entire population prior to an audit, the disallowed amounts are only known \textit{after} the audit and only for the \textit{sampled} claims. We will use both the terms `disallowed amount' and `error amount' to refer to the portion of a claim total that is not allowable for reimbursement.

\begin{table}
\centering
\begin{tabular}{ccccc}
\hline 
Line &Patient ID & Date of Service & Claimed Amount & Disallowed/Error Amount  \\ 
 &&&(known for & (only known \\
 &&& entire population) &for sampled claims)\\
\hline
1&33457 & Jan 15, 2017 & \$52.50 & \$0  \\ 
2&31415 & March 10, 2017 & \$78.90 & \$30.00 \\ 
3&44478 & Oct 27, 2016 & \$25.90 & \$25.90 \\
4&67841 & May 5, 2016 & \$105.00 & \$50.00 \\ 
5&55112 & Nov 20, 2016 & \$125.00 & \$0 \\ 
6&98765 & May 1, 2016 & \$66.00 & \$66.00 \\ 
\hline
\end{tabular} 
\caption{Portion of Hypothetical Data for a MediCal Audit}
 \label{auditDataTable}
\end{table}

The total disallowed amount found in the sample is extrapolated from the sample to the population, and the audited organization is required to pay that amount or some related amount back to the MediCal fund.  Clearly, maintaining a small margin of error in estimating the total disallowed amount is of interest to all parties. Thus, it is important to design audit samples which estimate the total disallowed amount with a reasonable margin of error while minimizing the sample size. In addition, since a pilot sample is typically an inconvenience to the organization being audited, audit samples must frequently be designed with little to no information about the population of disallowed amounts -- making it difficult to determine an appropriate sampling plan. \\

The text \textit{Statistical Auditing} by  \cite{roberts1978} likely contains the most comprehensive treatment of sample design issues for audit populations. In particular, Roberts derives estimates of the population variances needed for sample size determination under both simple expansion and ratio estimation under the all-or-nothing errors assumption.  His estimates do not require data from pilot samples.  However, they do require estimating the \textit{error rate}, defined as the proportion of claims in the population containing some error amount or disallowed amount. He uses a Bernoulli generative model to derive his estimates.    \cite{king2013} also propose a Bernoulli model to estimate the variance of the disallowed amounts under simple expansion under the all-or-nothing errors assumption but arrive at a slightly different estimate.  In this paper, we review and reconcile these two estimators.  In addition, since all currently available methods of determining sample size depend on estimating the error rate or the variance of the population of disallowed values, we also propose a method for determining a conservative sample size which is based solely on the claimed values and does not require any additional information about the population of disallowed values except the all-or-nothing errors assumption. 

Realistically, partial errors do occur in some audit populations so generalizing existing results and deriving new results that apply to more general error models is desirable.  We note that  \cite{neter1977} do consider more general error models in their empirical study, but our research has revealed little theoretic work on sample design under more general error models which can handle partial errors.  One exception is the penny sampling method proposed by \cite{edwards2015conservative}, which treats each penny in the total audited amount as a sampling unit and uses the inversion of a hypothesis test for a binomial proportion to obtain exact confidence intervals. One limitation of this work, however, is that it cannot be used for populations where underpayment to the MediCal provider is a possibility, i.e. penny sampling can only be used with populations where all errors are overpayments to the provider. Another exception is \cite{liu2005efficient} who consider a  partial-error model which assumes a quasi-uniform distribution of the partial error amount for each claim.  Liu et al. use this model to derive optimal strata breakpoints under ratio estimation in the audit setting.  However, the model of Liu et al. does not accurately model the line-item error mechanism which  generates  partial errors in healthcare audit populations.  Consequently, we develop a novel partial error model based on the underlying line-item errors and use it to extend Roberts' results on sample size for all-or-nothing error populations to audit populations with partial errors. Under our line-item error model, we additionally show that the resulting variance estimates, which depend on two possibly unknown error-rate parameters, can be maximized to obtain a conservative sample size for audits where estimates of the required error-rate parameters are not available.

We also consider the question of choosing between the simple expansion and ratio estimators in simple random sampling with all-or-nothing errors.  The general advice on p.157 of  \cite{cochran1977} is to use the ratio estimator instead of simple expansion when $X$, the claim amount, and $Y$, the disallowed amount, satisfy:
\begin{eqnarray}
\dfrac{\hbox{Cov}(X, Y)}{\sigma_x \sigma_y}&>& \dfrac{1}{2}\dfrac{(\dfrac{\sigma_x}{\mu_x})}{(\dfrac{\sigma_y}{\mu_y})} \label{ratiocriteria}
\end{eqnarray}
We specialize this inequality to the audit population setting, and derive a formula for the probability that ratio estimation will outperform simple expansion. Since our formula only relies on the error rate and parameters for the claim population, it can be used for sample planning prior to collecting any information about the population of disallowed amounts. 

We note that although we discuss our results in the context of MediCal/Medicaid audits, they are more generally applicable to any type of healthcare audit where 1) the sampling unit consists of an invoice which is composed of one or more line-item charges; 2) either the entire invoice amount or individual line-item amounts may be in error; and 3) line-item errors are either all-or-nothing errors or a pre-audit estimate of the amounts of all line-item partial errors is available.

\section{Notation and Estimators}
We now summarize some notation and the two common estimators used to extrapolate the total disallowed amount in an audit. 
\begin{eqnarray}
N &=& \hbox{the population size} \nonumber \\
\{x_1, x_2,...,x_N\}&=& \hbox{the population of known claimed amounts} \nonumber \\
\{y_1, y_2,...,y_N\} &=& \hbox{the population of unknown disallowed/error amounts} \nonumber \\
\tau_x &=& \sum_{i=1}^N x_i \nonumber \\
\tau_y &=& \sum_{i=1}^N y_i\nonumber \\
R &=& \dfrac{\tau_y}{\tau_x}\nonumber \\
\mu_x &=& \dfrac{1}{N} \tau_x \nonumber \\
\mu_y &=& \dfrac{1}{N}\tau_y \nonumber \\
\sigma_x^2 &=& \dfrac{1}{N}\sum_{i=1}^N (x_i-\mu_x)^2 \nonumber \\
\sigma_y^2 &=& \dfrac{1}{N}\sum_{i=1}^N (y_i-\mu_y)^2 \nonumber \\
\end{eqnarray}
\begin{eqnarray}
n &=& \hbox{the sample size}\nonumber \\
\{x_{(1)}, x_{(2)},...,x_{(n)}\} &=& \hbox{a sample random sample of claims without replacement} \nonumber \\
\{y_{(1)}, y_{(2)},...,y_{(n)}\} &=& \hbox{disallowed values corresponding to sampled claims} \nonumber \\
\bar{y}&=&\dfrac{1}{n}\sum_{i=1}^n y_{(i)} \nonumber \\
\bar{x}&=&\dfrac{1}{n}\sum_{i=1}^n x_{(i)} \nonumber \\
\hat{r} &=& \dfrac{\bar{y}}{\bar{x}}\nonumber \\
\hat{\tau}_{se}&=&N\bar{y}= \hbox{the simple expansion estimator of total error amount} \nonumber \\
\hat{\tau}_r&=&\hat{r} \tau_x = \hbox{the ratio estimator of total error amount} \nonumber\\
\sigma_{\hat{\tau}_{se}}^2 &=& N^2 \cdot \dfrac{\sigma_y^2}{n}\cdot \dfrac{N-n}{N-1} \label{varsimpexp} \\
\sigma_R^2 &=& \dfrac{1}{N} \sum_{i=1}^N(y_i-R x_i)^2 \nonumber \\
\sigma_{\hat{\tau}_{r}}^2 &=& N^2 \cdot \dfrac{\sigma_R^2}{n}\cdot \dfrac{N-n}{N-1}\nonumber
\end{eqnarray}

The rest of this paper is organized as follows.  In Section 3, we  give the sample size formula of interest.  In Section 4, we discuss issues pertaining to sample size determination under the simple expansion estimator.  In particular, we review the existing binomial generative model for audit populations; reconcile the estimators of the variance under all-or-nothing errors proposed by \cite{roberts1978} and \cite{king2013}; propose a new partial-error model and extend the procedure for estimating variance to the proposed partial error model; and maximize the variance under all-or-nothing or partial errors to obtain conservative sample sizes that do not require pilot study information.  In Section 5, we consider the ratio estimator.  We start with a criteria for deciding between simple expansion and ratio estimation; review the estimator for the variance of the ratio estimator under simple random sampling proposed by \cite{roberts1978}; extend this estimate of variance under the proposed partial-error model; maximize the variance and derive a procedure for calculating a conservative sample size; and finish with comments about optimal stratification under ratio estimation.  In Section 6, we apply the sample design tools developed in this paper to three simulated audit populations.  We offer some concluding remarks and avenues for further research in Section 7.

\section{Sample Size Formula}
Under simple expansion, the $(1-\alpha) \times 100 \%$ large-sample confidence level margin of error of $\hat{\tau}_{se}$ is
\begin{eqnarray}
E=z_{(1-\frac{\alpha}{2})}\cdot \sigma_{\hat{\tau}_{se}}\label{marginSimpExp}
\end{eqnarray}
where $z_p$ denotes the $p$th percentile of the standard normal distribution. 
Substituting equation (\ref{varsimpexp}) into equation (\ref{marginSimpExp}) and solving for $n$, we obtain the following sample size formula under simple expansion
\begin{eqnarray}
n=\dfrac{z_{(1-\frac{\alpha}{2})}^2 \cdot N^3\sigma_y^2}{E^2(N-1)+z_{(1-\frac{\alpha}{2})}^2N^2\sigma_y^2} \label{SampleSizeSimExp}
\end{eqnarray}
The sample size formula will give the sample size required to attain a chosen margin of error and confidence level provided that the variance of disallowed amounts, $\sigma_y^2$, is known.  However, $\sigma_y^2$ is typically not known in the planning stages of an audit. One could obtain an estimate of $\sigma_y^2$ using a pilot sample, but this is an inconvenience to an audited organization since they would have to pull records twice -- once for the pilot sample and again for the actual full audit. In the next section, we propose a generative model for audit populations which permits estimation of $\sigma_y^2$ in cases where the error rate can be approximated.  Under ratio estimation, the sample size formula is equation (\ref{SampleSizeSimExp}) with $\sigma_R^2$ substituted for $\sigma_y^2$.  We propose methods for estimating $\sigma_R^2$ during the planning stages of an audit in Section \ref{predictsigmar}. 


\section{The All-or-Nothing Errors Model}
\cite{roberts1978} and \cite{king2013} both formulate estimates of $\sigma_y^2$ for audit populations with all-or-nothing errors.  They assume the audit population was generated in such a way that the entire claim amount is in error with probability $\pi$ or none of the claim amount is in error with probability $1-\pi$.  They additionally assume errors are made independently from claim to claim.  More formally, letting $X_i=$ be the value $i$th claim in the population, $U_i$ be an error indicator variable, $Y_i=$ the error/disallowed value of the $i$th claim for $i=1,2,...,N$ and $0\leq \pi \leq 1$, they propose the audit population is generated as follows: 
\begin{eqnarray}
  U_i&=& \left\{
\begin{array}{ll}
      1 & \hbox{ with probability } \pi \\
      0 & \hbox{ with probability } 1- \pi \\
     
\end{array} 
\right. \label{modelBinom}\\
 \hbox{and}&&\nonumber\\
Y_i&=&U_iX_i\nonumber\end{eqnarray}

Although they use the same generative model, Roberts and King et al. propose  different estimates of $\sigma_y^2$. Roberts uses the expected value of the population variance of $Y_1, Y_2,...,Y_N$, where the expectation is taken over all potential audit populations.  We denote Robert's estimate as $E_{\vec{U}}(\sigma_y^2)$ where $\vec{U}=U_1, U_2,...,U_N$.  This estimate can be computed using the formula in equation (\ref{Robertsvar}).
\begin{eqnarray}
\hat{\sigma}_{(R,y)}^2 &=&E_{\vec{U}}(\sigma_y^2)\nonumber\\
&=& E_{\vec{U}}[\dfrac{1}{N}\sum_{i=1}^N Y_i^2 -\dfrac{1}{N^2}(\sum_{i=1}^N Y_i)^2]\nonumber\\
&=&\pi \mu_x^{(2)}-(\pi \mu_x)^2-\pi(1-\pi)\dfrac{\sigma_x^2+\mu_x^2}{N} \label{Robertsvar}
\end{eqnarray}
On the other hand, \cite{king2013} use the total variance of $Y$, where $Y$ is defined to be a random draw from the random vector $Y_1, Y_2,...,Y_N$ to estimate $\sigma_y^2$. Thus, $Y$ can be interpreted as a random draw from a random audit population. $\var(Y)$ can be found using iterated expectations as shown in the proof to \ref{totalexpvar}. 
\begin{theorem}[Total Expected Value and Variance]\label{totalexpvar}
Under the model for $Y$ given in equation (\ref{modelBinom}):

\begin{enumerate}
\item[a.] $E(Y)=\pi \mu_x$
\item[b.] $\hbox{Var}(Y) = \pi\mu_x^{(2)}-(\pi \mu_x)^2$ 
\end{enumerate}
\end{theorem}

\textbf{Proof:} \begin{enumerate}
\item[a.]\begin{eqnarray*}
E(Y)&=&EE(Y|U_1, U_2,...,U_N)\\
&=& E(\frac{1}{N}\sum_{i=1}^N Y_i | U_1, U_2,...,U_N)\\
&=& E(\frac{1}{N}\sum_{i=1}^N U_i X_i)\\
&=& \frac{1}{N}\sum_{i=1}^N X_i E(U_i)\\
&=& \pi \frac{1}{N}\sum_{i=1}^N X_i \\
&=&\pi \mu_x
\end{eqnarray*}

\item[b.] 
\begin{eqnarray}
\var(Y)&=& E[\var(Y|U_1, U_2,...,U_N)]+\var[E(Y|U_1, U_2,...,U_N)]\nonumber \\
&=& \hat{\sigma}_{(R,y)}^2 +\var(\frac{1}{N}\sum_{i=1}^N U_i X_i )\nonumber\\
&=& \hat{\sigma}_{(R,y)}^2 +\frac{1}{N^2}\sum_{i=1}^N X_i^2 \var(U_i) )\nonumber\\
&=& \hat{\sigma}_{(R,y)}^2 +\frac{1}{N^2}\sum_{i=1}^N X_i^2 \pi(1-\pi) )\nonumber\\
&=& \hat{\sigma}_{(R,y)}^2 +\frac{1}{N} \pi(1-\pi)\mu_x^{(2)} \nonumber\\
&=&[ \pi \mu_x^{(2)}-(\pi \mu_x)^2-\pi(1-\pi)\dfrac{\sigma_x^2+\mu_x^2}{N}]+\frac{1}{N} \pi(1-\pi)\mu_x^{(2)} \nonumber\\
&=&\pi \mu_x^{(2)}-\pi^2\mu_x^2 + [-\pi(1-\pi)\dfrac{\mu_x^{(2)}}{N}+\frac{1}{N} \pi(1-\pi)\mu_x^{(2)}] \nonumber\\
&=&\pi \mu_x^{(2)}-(\pi\mu_x)^2
\end{eqnarray}
\end{enumerate} 

The federal Office of Inspector General's RAT-STATS software  also uses the total variance of 
$Y$, $\var(Y)$, to estimate $\sigma_y^2$ (RAT-STATs Companion Manual, Rev 5/2010, p. 4-9). The total variance, however, represents the variation in $Y$ as the audit population and the sample from it vary.  We would argue, however, that the audit population is fixed but unknown so that including variation due to a varying audit population in our estimation of $\varyy$ is not conceptually satisfying.  In addition, since the Roberts estimate will minimize the mean square prediction error, we prefer it over the total variance.
 
The two proposed estimators of $\sigma_y^2$ are related by the following inequality:

\[\hat{\sigma}_{(R,y)}^2  \leq \var(Y)\] 
However, if the population size, $N$ is large relative to $\sigma_x^2$ and $\mu_x^2$, then the term $\pi(1-\pi)(\sigma_x^2+\mu_x^2)/N$ in $\hat{\sigma}_{(R,y)}^2$ will  be small relative to $\mu_x^{(2)}-(\pi \mu_x)^2$ so that

\[\var(Y)=\pi \mu_x^{(2)}-(\pi\mu_x)^2 \approx \pi \mu_x^{(2)}-(\pi \mu_x)^2-\pi(1-\pi)\dfrac{\sigma_x^2+\mu_x^2}{N}=\hat{\sigma}_{(R,y)}^2\]
i.e. the proposed estimators will be roughly equal.  This has been the case in several audit populations we have reviewed.

\subsubsection{Estimating $\pi$}\label{conservsmp}
The formula for $\hat{\sigma}_{(R,y)}^2$ in equation (\ref{Robertsvar}) only depends on the known population of claimed amounts and the error rate, $\pi$.  So we can use $\hat{\sigma}_{(R,y)}^2$ if an estimate of $\pi$ is available from a past survey or a pilot survey, then substitute the result into equation (\ref{SampleSizeSimExp}) to determine the sample size needed to achieve a given margin of error and confidence level.

If an estimate of $\pi$ is not available, we can obtain a conservative sample size by maximizing $h(\pi)=\hat{\sigma}_{(R,y)}^2=\pi \mu_x^{(2)}-(\pi \mu_x)^2-\pi(1-\pi)\mu_x^{(2)}/N$ as a function of $\pi$.  Taking the derivative of $h(\pi)$ and setting it equal to 0 gives:

\begin{eqnarray}
h'(\pi) = \mu_x^{(2)}-2\pi \mu_x^2-(1-2\pi)\dfrac{\mu_x^{(2)}}{N}=0 \label{derivOfh}
\end{eqnarray}
Solving equation (\ref{derivOfh}), we obtain
\begin{eqnarray}
\pi_{\hbox{crit}} = \dfrac{1}{2}\cdot \dfrac{\momtwo-\dfrac{\mu_x^{(2)}}{N}}{\mu_x^2-\dfrac{\mu_x^{(2)}}{N}}\approx \dfrac{\momtwo}{2 \mu_x^2} \label{pi.max}
\end{eqnarray}

In order to maximize $h(\pi)$ over $\pi \in [0,1]$, we must check $h(0), h(1)$ and $h(\pi_{\hbox{crit}})$.  Since $h(0)=0$ and $h(
1)=\sigma_x^2$, the maximum value of $h(\pi)$ is $h_{\hbox{max}}=\hbox{max}\{\sigma_x^2,h(\pi_{\hbox{crit}})\}$. The sample size obtained by substituting $h_{\hbox{max}}$ for $\varyy$ in equation (\ref{SampleSizeSimExp}) will be the maximum sample size needed for a specified margin of error and confidence level over all possible error rates, $\pi$.

\subsection{Partial Payments}
Thus far, we have considered a model with all-or-nothing errors. We now wish to consider sample size determination under simple expansion when there are partial payments in the population, i.e. only a portion of the amount claimed is deemed allowable and the remaining portion is disallowable. \cite{liu2005efficient} proposed the partial payment model in equation (\ref{liupartial}).

\begin{eqnarray}
  Y_i= \left\{
\begin{array}{ll}
      pX_i+u(1-p)X_i & \hbox{ with probability } p \\
      pX_i-upX_i & \hbox{ with probability } 1- p \\
     
\end{array} 
\right. \label{liupartial}
\end{eqnarray}
for $i=1,2,...,N$ and where $u \sim U(0,1)$ and $p$ is the proportion of claims in the population having an error. This model assumes a uniform distribution over all potential error amounts below the average partial error ($pX_i$) and a uniform distribution over all error amounts above the average partial error amount. However, in MediCal audits, the partial error amount of a claim typically arises from fixed, discrete amounts corresponding to errors in underlying line-item charges.   For example, Table \ref{lineitemtable} shows the detailed line-item charges for a single MediCal claim for a fictitious patient. The claim consists of three line items -- one for each billable service provided by the medical provider to the patient on his/her June 1, 2017 visit.
\vspace{.3cm}

\begin{tabular}{cccr}
\hline 
Patient ID & Date of Service & Procedure & Claimed amount \\ 
\hline
1234 & June 1, 2017 & Office Visit, Level 4 & \$45.00 \\ 
 
1234 & June 1, 2017 & Blood Test & 6.00 \\ 
 
1234 & June  1, 2017 & x-ray & 17.00 \\ 
\hline 
 &  & Total & \$68.00 \\ 
 \hline
 \label{lineitemtable}
\end{tabular} 
\begin{center}
Table \ref{lineitemtable}
\end{center}

All-or-nothing errors can occur for any line item.  It is also possible for a line item to be partially in error.  Partial line-item errors occur when a billed procedure is downgraded to a lower level of service.  For example, if MediCal was billed for a level 4 office visit, but documentation about the patient's condition does not substantiate a level 4 office visit (based on the complexity of the case) then the procedure may be downgraded by the auditor to a level 3 office visit. The amount reimbursable by MediCal will also be adjusted, say from \$45.00 to \$40.00,  resulting in a partial error of \$5.00 for that line item. 

We propose a partial error model which models the error/disallowed amount of a claim as the sum of the line-item disallowed amounts in that claim.  We further assume that errors occur independently from line to line with the same probability $\pi_L$ on each line.  In order to define the line-item model, we introduce some notation:

\begin{eqnarray*}
b_i &=& \hbox{the number of lines in claim $i$ for $i=1,2,...,N$}\\
X_{ij}&=& \hbox{the claimed amount for line $j$ of claim $i$}\\
Y_{ij}&=& \hbox{the error/disallowed amount for line $j$ of claim $i$}\\
\tilde{X}_{ij} &=& \hbox{the most probable error amount for line $j$ of claim $i$}\\
\tilde{X}_i &=& \hbox{the sum of the most probable error amounts for claim $i$}\\
\pi_L&=& \hbox{the probability of a line-item error}
\end{eqnarray*}

The most probable error amount, $\tilde{X}_{ij}$, will be $X_{ij}$ for all-or-nothing line items and may be taken as the amount associated with one level of service below that which was claimed for downgradable line items (unless some auxiliary information suggests a better alternative).  We can express the proposed partial error model as follows:

\begin{eqnarray}
  Y_{ij}= \left\{
\begin{array}{ll}
      \tilde{X}_{ij} & \hbox{ with probability } \pi_L \\
      0 & \hbox{ with probability } 1- \pi_L \\
     
\end{array} 
\right. \label{modelBinomline}
\end{eqnarray}
Letting $W_{ij} \sim \hbox{Bern}(\pi_L)$ be a line-item error indicator variable and recalling that $U_i \sim \hbox{Bern}(\pi)$ is the claim-level error indicator,
the claim level error/disallowed amount can be expressed as the sum of a term representing the entire amount of the claim for a claim-level error plus the sum of the line item errors if there is no claim-level error as shown in equation (\ref{auditamtVarPartialErrors}). 

\begin{eqnarray}
Y_i &=& U_i X_i +(1-U_i)\sum_{j=1}^{b_i}Y_{ij}\nonumber\\
&=&  U_i X_i +(1-U_i)\sum_{j=1}^{b_i}W_{ij} \tilde{X}_{ij} 
\label{auditamtVarPartialErrors}
\end{eqnarray}
Using this model, we extend Roberts' estimate of $\sigma_y^2$ under all-or-nothing errors to allow for partial errors. Letting $\vec{W}=\{W_{ij}: i=1,2,...,N \hbox{ and } j=1,2,...,b_i\}$, we propose $E_{\vec{U}, \vec{W}}(\sigma_y^2 )$ as an estimate of $\sigma_y^2$.  We assume that the vectors of claim-level and line-item error indicator variables are independent, i.e. $\vec{U} \perp \vec{W}$.

\begin{eqnarray}
E_{\vec{U}, \vec{W}}(\sigma_y^2 )&=&E_{\vec{U}, \vec{W}}(\frac{1}{N}\sum_{i=1}^N Y_i^2-(\frac{1}{N}\sum Y_i)^2 ) \nonumber\\
&=& \frac{1}{N}\sum_{i=1}^N E(Y_i^2)-\frac{1}{N^2}E((\sum Y_i)^2 ) \nonumber\\
&=& \frac{1}{N}\sum_{i=1}^N E(Y_i^2)-\frac{1}{N^2}E(\sum Y_i^2+\sum_{i=1}^N\sum_{i'\neq i, i'=1}^N Y_iY_{i'})\nonumber\\
&=&(\frac{1}{N}-\frac{1}{N^2})\sum_{i=1}^N E(Y_i^2)-\frac{1}{N^2}\sum_{i=1}^N \sum_{i'\neq i, i'=1}^N E(Y_iY_{i'}) \label{partialvary}
\end{eqnarray}
We now derive $E(Y_i^2)$ and $E( Y_i Y_{i'} )$ to substitute back into equation (\ref{partialvary}). 

\begin{eqnarray}
E(Y_i^2)&=& E[( U_i X_i +(1-U_i)\sum_{j=1}^{b_i}W_{ij} \tilde{X}_{ij} )^2]\nonumber\\
&=& [\pi X_i^2+(1-\pi)\pi_L^2\tilde{X}_i^2]+(1-\pi)\pi_L(1-\pi_L)\sum_{j=1}^{b_i}\tilde{X}_{ij}^2 \label{secMoment}
\end{eqnarray}

\begin{eqnarray}
E( Y_i Y_{i'} )&=&E[(U_i X_i +(1-U_i)\sum_{j=1}^{b_i}W_{ij} \tilde{X}_{ij} )\cdot(U_{i'} X_{i'} +(1-U_{i'})\sum_{j=1}^{b_{i'}}W_{i'j} \tilde{X}_{i'j} )]\nonumber\\
&=& \pi^2X_iX_{i'}+\pi(1-\pi)\pi_L(X_{i'}\tilde{X_i}+X_i\tilde{X_{i'}})+(1-\pi)^2\pi_L^2\tilde{X_i}\tilde{X_{i'}}\label{covPartial}
\end{eqnarray}

Substituting (\ref{secMoment}) and (\ref{covPartial}) into (\ref{partialvary}) and simplifying gives:

\begin{eqnarray}
E(\sigma_y^2) &=& (\frac{1}{N}-\frac{1}{N^2})[\pi \sum_{i=1}^N X_i^2+(1-\pi)\pi_L^2\sum_{i=1}^N \tilde{X}_i^2+(1-\pi)\pi_L(1-\pi_L)\sum_{i=1}^N\sum_{j=1}^{b_i}\tilde{X}_{ij}^2] \nonumber\\
&&-\frac{1}{N^2}[\pi^2\sum_{i=1}^N\sum_{i'\neq i, i'=1}^N X_iX_{i'}+\pi(1-\pi)\pi_L\sum_{i=1}^N\sum_{i'\neq i, i'=1}^N(X_{i'}\tilde{X_i}+ X_i\tilde{X_{i'}})\nonumber\\
&&+(1-\pi)^2\pi_L^2\sum_{i=1}^N\sum_{i'\neq i, i'=1}^N\tilde{X_i}\tilde{X_{i'}}]\label{partialvaryformula}
\end{eqnarray}

In the case where line item errors are all-or-nothing, $\tilde{X}_{ij}=X_{ij}$ for all $i$ and $j$ so equation (\ref{partialvaryformula}) simplifies to:

\begin{eqnarray}
E(\sigma_y^2) &=& (\frac{1}{N}-\frac{1}{N^2})[(\pi +(1-\pi)\pi_L^2)\sum_{i=1}^N X_i^2+(1-\pi)\pi_L(1-\pi_L)\sum_{i=1}^N\sum_{j=1}^{b_i}X_{ij}^2] \nonumber\\
&&-\frac{1}{N^2}[\pi +(1-\pi)\pi_L]^2[(\sum_{i=1}^N X_i)^2-\sum_{i=1}^N X_i^2]\nonumber\\
\end{eqnarray}
We note that all quantities in equation (\ref{partialvaryformula}) are known from the claim data available prior to the audit except the claim-level error rate, $\pi$, and the line-item error rate, $\pi_L$.  Thus, equation (\ref{partialvaryformula}) can be used to estimate $\varyy$ if estimates of $\pi$ and $\pi_L$ are available from past surveys or pilot study data. We address situations where estimates of these two parameters are not available in the next section. 

\subsection{Conservative Sample Size}
Since $E(\sigma_y^2)$ is a polynomial in $\pi$ and $\pi_L$, we can maximize $E(\sigma_y^2)$ over $(\pi, \pi_L)$ to determine a conservative sample size which will be sufficient for any combination of $(\pi, \pi_L) \in [0,1] \times [0,1]$. To simplify notation, we define:

\begin{eqnarray}
c_1&=& (\frac{1}{N}-\frac{1}{N^2}) \sum_{i=1}^N X_i^2 \nonumber\\
c_2&=& (\frac{1}{N}-\frac{1}{N^2})\sum_{i=1}^N \tilde{X}_i^2 \nonumber\\
c_3 &=& (\frac{1}{N}-\frac{1}{N^2})\sum_{i=1}^N\sum_{j=1}^{b_i}\tilde{X}_{ij}^2 \nonumber\\
c_4 &=& -\frac{1}{N^2} \sum_{i=1}^N\sum_{i'\neq i, i'=1}^N X_iX_{i'} \nonumber\\
c_5 &=&- \frac{1}{N^2}\sum_{i=1}^N\sum_{i'\neq i, i'=1}^N(X_{i'}\tilde{X_i}+ X_i\tilde{X_{i'}}) \nonumber\\
c_6 &=& -\frac{1}{N^2}\sum_{i=1}^N\sum_{i'\neq i, i'=1}^N\tilde{X_i}\tilde{X_{i'}} \nonumber
\end{eqnarray}

then the formula for $E(\sigma_y^2)$ given in equation (\ref{partialvaryformula}) can be written:

\begin{eqnarray}
E(\sigma_y^2) &=& h(\pi, \pi_L)=c_1\pi +c_2(1-\pi)\pi_L^2+c_3(1-\pi)\pi_L(1-\pi_L)+c_4\pi^2+c_5 \pi(1-\pi)\pi_L \nonumber \\
&&+c_6(1-\pi)^2\pi_L^2 \label{expVarY}
\end{eqnarray}

Taking the partial derivatives of $h$, we obtain:

\begin{eqnarray}
\dfrac{\partial h}{\partial \pi}&=&c_1-c_2 \pi_L^2-c_3\pi_L(1-\pi_L)+2c_4 \pi+c_5(1-2\pi)\pi_L-2c_6(1-\pi)\pi_L^2 \nonumber \\
\dfrac{\partial h}{\partial \pi_L} &=& (1-\pi)[2c_2\pi_L+c_3(1-2\pi_L)+c_5\pi+2c_6(1-\pi)\pi_L] \label{partialsvary}
\end{eqnarray}

Setting the partial derivatives equal to 0 and solving for $\pi_L$ results in the following cubic equation 

\begin{eqnarray}
&-2c_6(c_2-c_3)\pi_L^3+3c_5(c_2-c_3)\pi_L^2+[-4c^*c_4+c_3c_5+c_5^2-2c_1c_6]\pi_L-2c_3c_4+c_1c_5=0\nonumber\\
&\hbox{where } c^*=c_2-c_3+c_6 
\end{eqnarray}
Thus, setting the partials equal to 0 will yield at most three critical values of $h(\pi, \pi_L)$.  We also check for possible maximums on the boundaries $\pi=0, \pi=1, \pi_L=0$ and $\pi_L=1$ by separately maximizing  equations (\ref{firsteq})-(\ref{lasteq}). 

\begin{eqnarray}
h(0, \pi_L)&=& c_2\pi_L^2+c_3\pi_L(1-\pi_L)+c_6\pi_L^2 \label{firsteq}\\
h(1, \pi_L)&=&c_1+c_4 = \sigma_x^2  \\
h(\pi,0)&=& c_1\pi+c_4\pi^2 = \hat{\sigma}_{(R,y)}\label{boundary}\\
h(\pi,1) &=& c_1\pi+c_2(1-\pi)+c_4\pi^2+c_5\pi(1-\pi)+c_6(1-\pi)^2 \label{lasteq}
\end{eqnarray}
Examining the boundary equations, we observe that $h$ is either a constant or a quadratic function on each boundary and, hence, is easily maximized on any boundary. The conservative sample size is determined by taking $h(\pi,\pi_L)=E(\sigma_y^2)$ to be its maximum value over any real-valued critical points that fall in $(0,1) \times (0,1)$ and over the maxima from the four boundaries.

\section{Ratio Estimation}
In this section, we show that, in the all-or-nothing errors case, ratio estimation is expected to outperform simple expansion for any audit population, provided the assumptions are met for the use of ratio estimation. We then review the estimator of the variance, $\sigma_R^2$, needed under ratio estimation which was proposed in \cite{roberts1978} under the all-or-nothing error assumption.  This proposed estimator of $\sigma_R^2$ depends on the error rate $\pi$, and we observe that $\pi=0.50$ maximizes the estimated value of $\sigma_R^2$. Thus, in cases where $\pi$ is unknown, a conservative sample size can be computed in the all-or-nothing errors case. We comment on stratification under ratio estimation. Finally,
we derive an estimate of the variance for the line-item partial errors model and show that a conservative sample size can be computed under this model. 
\subsection{Choosing Between Ratio Estimation and Simple Expansion}
We now derive a method for determining whether ratio estimation or simple expansion will be more efficient for extrapolating data from an audit sample. Rearranging the criteria (in inequality (\ref{ratiocriteria})) for choosing between these two estimators gives
\begin{eqnarray}
\cov(X, Y)-\dfrac{\varx}{2\mu_x}\mu_y>0 \label{ratiocriteria2}
\end{eqnarray}

Under the binomial generative model, we have
$\hbox{Cov}(X, Y|\pi) = E(XY) -\mu_x \mu_y=(1/N)\sum_{i=1}^N X_i Y_i-\mu_x \mu_y$ and $\mu_y=(1/N)\sum_{i=1}^NY_i$ . Making these substitutions into inequality (\ref{ratiocriteria2}) and simplifying, we obtain:
\begin{eqnarray}
 g(\vec{U})&=&\cov(X, Y)-\dfrac{\varx}{2\mu_x}\mu_y\nonumber
 \\
&=& [\frac{1}{N}\sum_{i=1}^N X_i Y_i - \mu_x \cdot \frac{1}{N} \sum_{i=1}^N Y_i] - \dfrac{\varx}{2\mu_x}\cdot \frac{1}{N} \sum_{i=1}^N Y_i\nonumber \\
&=& \frac{1}{N}\sum_{i=1}^N Y_i(X_i-\mu_x-\dfrac{\varx}{2\mu_x})\nonumber \\
&=& \frac{1}{N}\sum_{i=1}^N U_iX_i(X_i-\mu_x-\dfrac{\varx}{2\mu_x})\nonumber \\
&=& \frac{1}{N}\sum_{i=1}^N c_i U_i
 \label{ratiocriteria4}\\
 &\hbox{where}\nonumber\\
 c_i&=& X_i(X_i-\mu_x-\dfrac{\varx}{2\mu_x}) \nonumber
\end{eqnarray}
The probability that $g(\vec{U})>0$ will represent our confidence that the ratio estimator will have smaller variance than the simple expansion estimator.  In order to compute this probability, we determine the distribution of $g(\vec{U})$. Often MediCal claim data consist of only a few distinct values, each of which is repeated a large number of times.  Suppose there are $v$ distinct claim total values, $X_{(1)}, X_{(2)},...,X_{(v)}$ resulting in the corresponding $v$ distinct values of $c_i$,  $c_{(1)}, c_{(2)},...,c_{(v)}$.  Let $S_l$ for $l=1,2,...,v$ be the set of subscripts of claims having the value $X_{(l)}$ and $N_l$ be the number of elements in $S_l$.  Then the criteria for choosing between ratio estimation and simple expansion becomes:

\begin{eqnarray}
 g(\vec{U})&=& \frac{1}{N}\sum_{k=1}^N c_i U_i \nonumber\\
&=& \frac{1}{N} \sum_{l=1}^v \sum_{i\in S_l} c_{(l)} U_i\nonumber \\
&=&  \sum_{l=1}^v \frac{c_{(l)}}{N}\sum_{i\in S_l}  U_i 
\label{criteria}
\end{eqnarray}
Recall the $U_i$ are independent and identically distributed Bernoulli random variables.  Thus, for each $l$, the summation $\sum_{i\in S_l}  U_i$ will be approximately normally distributed by the Central Limit Theorem if $|S_l|=N_l$ is large.  In this case, $g(\vec{U})$ will be approximately normally distributed since it is a linear combination of the approximately normal and independent random variables $\sum_{i\in S_l}  U_i$.
Additionally,  using linear operator properties of the mean and variance, the mean and variance of $g(\vec{U})$ can be shown to be:
\begin{eqnarray}
E(g(\vec{U}))&=& \dfrac{\pi}{2}\varx \nonumber\\
\var(g(\vec{U})) &=&\dfrac{\pi(1-\pi)}{N^2}\sum_{i=1}^N c_i^2\label{varofg}\\
\end{eqnarray}

Let  $Z$ represent the standard normal variate. If $N_l$ is large for all $l$,
\begin{eqnarray}
P(g(\vec{U})>0)
&\approx& P(Z > \dfrac{0-\dfrac{\pi}{2}\varx}{\sqrt{\dfrac{\pi(1-\pi)}{N^2}\sum_{i=1}^N c_i^2}})\nonumber\\
 &=& P(Z > \dfrac{-\dfrac{1}{2}\varx}{\sqrt{(\frac{1}{\pi}-1)\sum_{i=1}^N \frac{c_i^2}{N^2}}})\label{negnumerator}\\ 
&>& \dfrac{1}{2}\nonumber
\end{eqnarray}
where the last line is true since the numerator of the right side of line (\ref{negnumerator}) is negative.  The last line implies that ratio estimation is always favored to outperform simple expansion in \textit{any} claim population provided we can assume $g(\vec{U})$ is approximately normally distributed. Examining equation (\ref{negnumerator}), we see that as $\pi \to 0$, the probability ratio estimation is preferred approaches 0.5, and as $\pi \to 1$, the probability that ratio estimation is preferred approaches 1. If normality of $g(\vec{U})$ is not reasonable, a Monte Carlo estimate of the probability that inequality (\ref{ratiocriteria2}) is true would give a more accurate estimate of our confidence that ratio estimation will outperform simple expansion.

As noted by \cite{neter1977} and \cite{edwards2011}, ratio-estimator-based confidence intervals can fail to attain the nominal confidence level when applied to audit populations even if the standard large-sample criteria for using ratio estimation are met.  The excess zeros and skewness often found in audit populations require one to check normality assumptions under either estimator to ensure nominal confidence levels are likely to be met with the proposed sample size.  This can be done through Monte Carlo simulation under a range of potential error rates prior to starting an audit.

\subsection{Estimating $\sigma_R^2$} \label{predictsigmar}
Assuming all-or-nothing errors, \cite{roberts1978} proposes the following estimator of $\sigma_R^2$ under the binomial generative model:
\begin{eqnarray}
E(\varr | \pi)=\pi(1-\pi)\mu_x^{(2)}[1+\dfrac{1}{N}(\dfrac{\sigma_x^2}{\mu_x^2}+\dfrac{4}{1+(\frac{\sigma_x}{\mu_x})^2}-\dfrac{G_1}{\frac{\mu_x}{\sigma_x}[1+(\frac{\mu_x}{\sigma_x})^2]}-5)]\label{robertsratio}\\
\hbox{where } G_1=\dfrac{\sum_{j=1}^N(X_j-\mu_x)^3}{N\sigma^3} \nonumber
\end{eqnarray}

For large values of $N$, equation (\ref{robertsratio}) may be simplified to $E(\varr | \pi)  \approx \pi(1-\pi)\mu_x^{(2)}$. Note that equation (\ref{robertsratio}) depends on the error rate $\pi$.  As in the case of estimating a binomial proportion, (\ref{robertsratio}) is maximized if $\pi = 1/2$.  If the error rate cannot be estimated beforehand, using  $\pi = 1/2$ will yield a sample size which is sufficiently large for any value of $\pi$ when ratio estimation is to be used.

A problem which may be encountered when using ratio estimation for audit data is that the sample data may contain no errors.  In this case, the sample estimate of $\varr$ is $\hat{\sigma}_R^2 = \sum_{i=1}^n (Y_i-\hat{R}X_i)^2/(n-1)=\sum_{i=1}^n (0-0\cdot X_i)^2/(n-1)=0$ since $Y_i=0$ for $i=1,2,...,n$ and $\hat{R}=(\sum_{i=1}^n Y_i)/(\sum_{i=1}^n X_i)=0/(\sum_{i=1}^n X_i)=0$. Thus, the estimated margin of error is zero.  This problem can be resolved by obtaining an exact 90\% or 95\% lower bound for $\pi$ using the sample data and substituting it into (\ref{robertsratio}) to obtain a conservative estimate of $\varr$.

\subsection{Stratification Under Ratio Estimation}
Using the following notation, 
\begin{eqnarray*}
h &=& 1,2,...,L \hbox{ denotes the stratum}\\
N_h &=& \hbox{ the number of population units in stratum } h\\
x_{1h}, x_{2h},...,x_{Nh,h} &=& \hbox{ denotes the claim amounts in stratum } h\\
y_{1h}, y_{2h},...,y_{Nh,h} &=& \hbox{ denotes the error/disallowed amounts in stratum } h\\
n_h &=& \hbox{ the sample size for stratum } h\\
s_h \subseteq  \{1,2,...,N_h\}&& \hbox{ denotes the subscripts of claims in the sample from stratum } h \\
\mu_{yh}&=& \dfrac{1}{N_h}\sum_{i=1}^{N_h} y_{ih}\\
\sigma_{yh}^2&=&\dfrac{1}{N_h}\sum_{i=1}^{N_h}(y_{ih}-\mu_{yh})^2\\
\bar{Y_h}&=&\dfrac{1}{n_h} \sum_{i \in s_h} y_{ih}\\
\bar{X_h}&=&\dfrac{1}{n_h} \sum_{i \in s_h} x_{ih}\\
\tau_{xh} &=& \sum_{i=1}^{N_h} x_{ih}\\
\hat{R}_h &=& \dfrac{\bar{Y_h}}{\bar{X_h}}
\end{eqnarray*}
the stratified simple expansion and ratio estimators of $\tau_y$ and their variances are:
\begin{eqnarray}
\hat{\tau}_{simp,st} &=& \sum_{h=1}^{L}N_h \bar{Y_h} \nonumber\\
\hbox{Var}(\hat{\tau}_{simp,st}) &=& \sum_{h=1}^{L}  N_h^2 \cdot \dfrac{\sigma_{yh}^2} {n_h}\cdot \dfrac{N_h-n_h}{N_h-1}\label{varTauHat} \\
\hat{\tau}_{Ratio,st} &=& \sum_{h=1}^{L} \hat{R}_h \tau_{xh}\nonumber \\
\hbox{Var}(\hat{\tau}_{Ratio,st}) &=& \sum_{h=1}^{L}  N_h^2 \cdot \dfrac{\sigma_{Rh}^2} {n_h}\cdot \dfrac{N_h-n_h}{N_h-1} \label{varRatiostrat}
\end{eqnarray}
where $\sigma_{Rh}^2=\sum_{i=1}^{n_h}(y_{ih}-\hat{R}_h x_{ih})^2/(n_h-1)$.
Optimal stratification under simple expansion is found by choosing breakpoints that minimize Formula (\ref{varTauHat}). If $\sigma_{yh}^2$ is estimated by the Roberts' estimator in equation (\ref{Robertsvar}), the optimal stratification depends on the error rate, $\pi$, since $\pi$ cannot be factored out of the estimated values of $\sigma_{yh}^2$. The analogous problem of finding optimal strata breakpoints under ratio estimation is independent of $\pi$ since formula (\ref{varRatiostrat}) with $\sigma_{Rh}$ estimated by formula (\ref{robertsratio})  only includes $\pi$ in the multiplicative constant $\pi(1-\pi)$, which is the same in every term and can be factored out of the summation.  Thus, the optimal strata breakpoints under ratio estimation will be correct even if the estimated value of $\pi$ is incorrect.  Under simple expansion, however, incorrect estimation of $\pi$ can lead a suboptimal choice of strata breakpoints.

\subsection{Partial Errors under Ratio Estimation}
Under the line-item partial error model defined in (\ref{modelBinomline}), we now derive the expected value of $\varr$.

\begin{eqnarray}
E(N\varr) &=& E(\sum_{i=1}^N(Y_i-RX_i)^2)\nonumber\\
&=& E(\sum_{i=1}^N Y_i^2 -2R\sum_{i=1}^N X_i Y_i +R^2\sum_{i=1}^N X_i^2)\nonumber\\
&=& E(\sum_{i=1}^N Y_i^2 -2(\dfrac{\sum_{i=1}^N{Y_i}}{\sum_{i=1}^N{X_i}})\sum_{i=1}^N X_i Y_i +(\dfrac{\sum_{i=1}^N{Y_i}}{\sum_{i=1}^N{X_i}})^2\sum_{i=1}^N X_i^2)\nonumber 
\end{eqnarray}
After some algebra, we obtain:

\begin{eqnarray}
E(N\varr)= \sum_{i=1}^N(1-\frac{2}{\tau_x} X_i+\dfrac{\tau_x^{(2)}}{\tau_x^2})E(Y_i^2)+\sum_{i=1}^N \sum_{i'\neq i, i'=1}^N( -\frac{2}{\tau_x} X_i+\dfrac{\tau_x^{(2)}}{\tau_x^2})E(Y_iY_{i'}) \label{sigmarpartial}
\end{eqnarray}
Letting $\displaystyle{k_i =-\frac{2}{\tau_x} X_i+\dfrac{\tau_x^{(2)}}{\tau_x^2}}$ and substituting the formulas for $E(Y_i^2)$ and $E(Y_i Y_{i'})$ from (\ref{secMoment}) and (\ref{covPartial}) into equation (\ref{sigmarpartial}), we obtain:

\begin{eqnarray}
E(\varr)&=& a_1 \pi (1-\pi)+a_2(1-\pi)\pi_L^2+a_3(1-\pi)\pi_L(1-\pi_L)+a_4 \pi(1-\pi)\pi_L
\nonumber\\
&&+a_5(1-\pi)^2\pi_L^2 \label{ratiovarline} \\
\hbox{where} \nonumber\\
a_1 &=& \frac{1}{N}\sum_{i=1}^N(1+k_i)X_i^2 \nonumber \\
a_2 &=& \frac{1}{N} \sum_{i=1}^N(1+k_i)\tilde{X}_i^2 \nonumber \\
a_3 &=& \frac{1}{N} \sum_{i=1}^N[(1+k_i)\sum_{j=1}^{b_i}\tilde{X}_{ij}^2] \nonumber \\
a_4 &=&  \frac{1}{N}\sum_{i=1}^N \sum_{i'\neq i, i'=1}^N k_i(X_{i'}\tilde{X}_i+X_i \tilde{X}_{i'}) \nonumber \\
a_5 &=& \frac{1}{N} \sum_{i=1}^N \sum_{i'\neq i, i'=1}^N k_i\tilde{X}_{i'}\tilde{X}_i \nonumber 
\end{eqnarray}  
As a check, it can be verified that equation (\ref{ratiovarline}) does reduce to the formula for the expected value of $\varr$ under the all-or-nothing error assumption in equation (\ref{robertsratio}) when there are no line-item errors, i.e. when $\pi_L=0$. 

As in the case of simple expansion under the line-item partial errors model, we can find the global maximum of $E(\varr)=g(\pi,\pi_L)$ by: 1) setting the partial derivatives equal to zero and solving the resulting system of equations and 2) checking for maxima on the boundaries.  Three of the boundaries need to be checked when maximizing $E(\varr)$ since $E(\varr)=g(1,\pi_L)=0$. The remaining boundaries are: 

\begin{eqnarray}
g(\pi,1)&=& (-a_1-a_4+a_5)\pi^2+(a_1-a_2+a_4-2a_5)\pi+(a_2+a_5)\nonumber \\
g(0, \pi_L)&=& (a_2-a_3+a_5)\pi_L^2+a_3\pi_L \nonumber\\
g(\pi, 0) &=& a_1 \pi(1-\pi) \label{boundaryVarR}
\end{eqnarray}
The quadratics in equations (\ref{boundaryVarR}) are straightforward to optimize once the coefficients have been calculated.

We will calculate and compare the partial error variance functions under simple expansion and ratio estimation for a simulated audit population in Section \ref{examplesection} and determine maximal values of $E(\sigma_y^2)$ and $E(\varr)$ over all possible values of $(\pi, \pi_L)$.

\section{Audit Example}\label{examplesection}
\begin{figure}
\centering
\includegraphics[scale=.9]{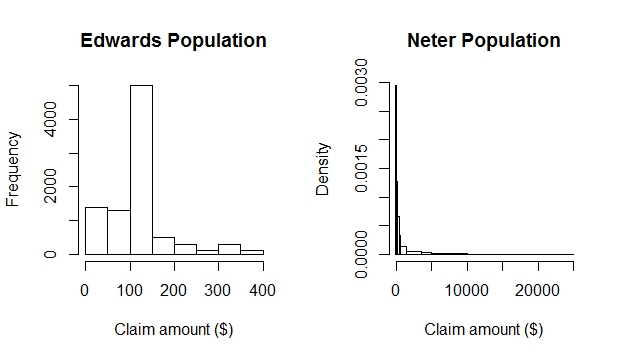}
\caption{The Two Simulated Audit Populations}\label{SimPopfig}
\end{figure}

Since actual MediCal audit data are confidential, we demonstrate these sample design tools using simulated audit populations. For populations with all-or-nothing errors, we simulate two populations. The \textbf{Edwards Population} was simulated to resemble the home health services population in  \cite{edwards2011}.  This population has a low variance and is right skewed with a spike of values in the \$100-150 range.  The population size is 9000, and it represents a paid amount of about \$1.1 million. The \textbf{Neter Population} was simulated to resemble Population 4 on p. 502 of \cite{neter1977}.  This population is also right skewed but with higher variance than the Edwards population.  It contains 4033 items and represents \$7.5 million. Histograms of these populations are shown in Figure  \ref{SimPopfig}.
For sample size determination in the presence of partial errors, we simulate a third population which is described in Section \ref{PartialExample}.

\subsection{All-or-nothing Errors Sample Size}
First we consider the determination of sample size under the all-or-nothing errors assumption.  

\subsubsection{Ratio Estimation versus Simple Expansion}
\begin{figure}
\centering
\includegraphics[scale=.8]{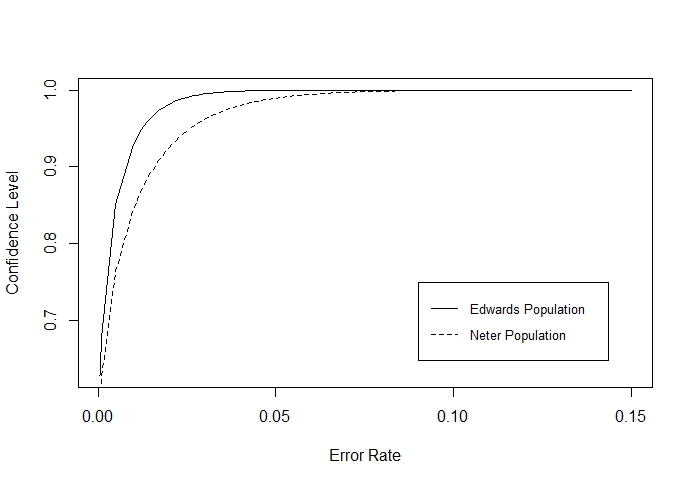} 
\caption{Confidence that Ratio Estimation will Outperform Simple Expansion}\label{ratiovssimplegraph}
\end{figure}

Using the criteria in inequality (\ref{negnumerator}), we can calculate the confidence that ratio estimation will outperform simple expansion over a range of potential error rates.  Figure \ref{ratiovssimplegraph} shows the results of this calculation with a separate graph for each population. Unless error rates are quite low, ratio estimation should be used for either population, assuming the assumptions for ratio estimation hold.

\subsubsection{All-or-nothing Errors:  Sample Size Under Ratio Estimation}
\begin{figure}
\centering
\includegraphics[scale=.8]{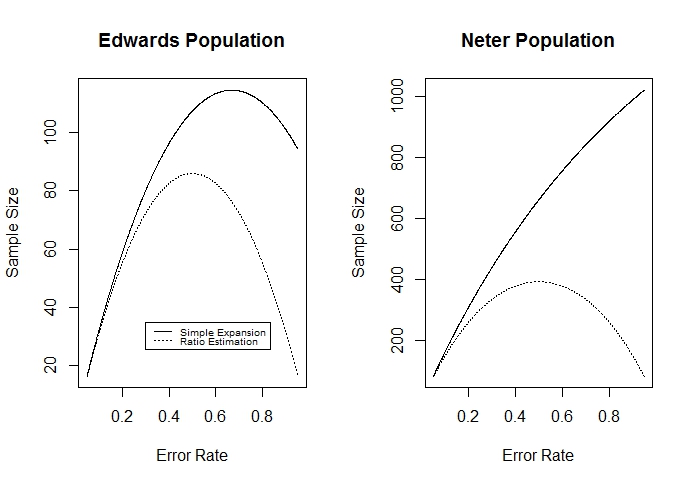}  
\caption{Sample Size as a Function of Error Rate}\label{samplesizefig}
\end{figure}

Suppose that for the Edwards Population representing \$1.1 million in paid claims, we wish to estimate the total error with maximum margin of error \$110,000 (10\% of the total amount paid) at 90\% confidence level.  For the Neter Population, representing \$7.5 million, we wish to estimate the total error with maximum margin of error \$750,000 at 90\% confidence. The sample sizes required over a range of potential error rates is shown for each population in Figure \ref{samplesizefig}. For comparison, the sample size is shown for both estimators even though the ratio estimator is the preferred estimator.
Under ratio estimation, maximal sample sizes occur at an error rate of $\frac{1}{2}$ for any population. This behavior is apparent in Figure \ref{samplesizefig}. However, under simple expansion, the error rate at which the maximal sample size occurs depends on the claim population data.  For the Edwards Population, the error rate at which the maximal sample size occurs is 0.67 using formula (\ref{pi.max}).  For the Neter Population, the error rate yielding maximal sample size is 2.72 which is outside the range of error rates, so $\pi=1$ will give the conservative sample size. 

\subsection{Sample Size Determination for the Simulated Partial Errors Population}\label{PartialExample}

\begin{figure}
\centering
\includegraphics[scale=1]{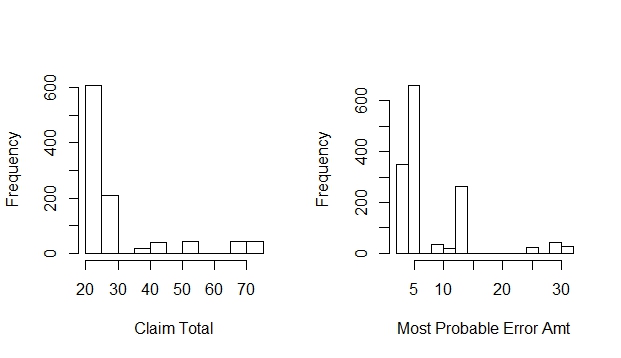} 
\caption{Claim Totals and Most Probable Line-item Error Amount for the Simulated Partial Error Audit Population}\label{graphPartialPop}
\end{figure}

The simulated population with partial errors, which we term the \textbf{Clinic Population}, consists of claims for 1000 patient-visits to a medical clinic. About 63\% of the claims have one line item, 33\% have two line items, and 4\% have three line items.  Histograms of the 1000 claim total amounts, $X_i$, and the 1416 ``most probable error amounts", $\tilde{X}_{ij}$, for each line item are shown in Figure \ref{graphPartialPop}. 

The mean and standard deviation of the claim totals are \$30.54 and \$13.43, respectively.  The mean and standard deviation of the most probable line-item error amounts are \$8.54 and \$6.45.  The total claimed amount is about \$30,500 for this population.

\begin{figure}
\centering
\includegraphics[scale=1]{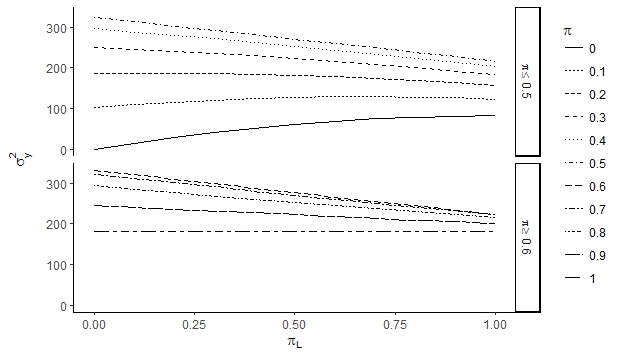} 
\caption{Cross-sections of $E(\sigma_y^2)=h(\pi, \pi_L)$ for fixed values of $\pi$}\label{graphVaryCrossSections}
\end{figure}

A graph of cross-sections of $E(\sigma_y^2)=h(\pi, \pi_L)$ in equation (\ref{expVarY}) for $\pi =0, 0.1, 0.2,...,1.0$ is shown in Figure \ref{graphVaryCrossSections}.  The only real-valued critical point of $h$ requires $\pi_L=2.58$ which is outside the domain of $h$.  Thus, $h$ will be maximized on its boundary; the boundaries of $h$ and the maximum value of $h$ on each is shown in Table \ref{boundarysimple}. Thus, the maximum value of $h$ is 306.\\
\vspace{.5cm}

\begin{table}
\centering
\begin{tabular}{cc}
\hline 
Boundary & Max of $h(\pi,\pi_L)=E(\sigma_y^2)$\\ 
\hline 
$\pi=0$ & 66 \\ 
$\pi=1$ & 154 \\ 
$\pi_L=0$ & 306 \\ 
$\pi_L=1$ & 199 
\end{tabular}
\caption{Maximum values of $E(\sigma_y^2)$ on Domain Boundaries}
 \label{boundarysimple}
\end{table}

\begin{figure}
\centering
\includegraphics[scale=1]{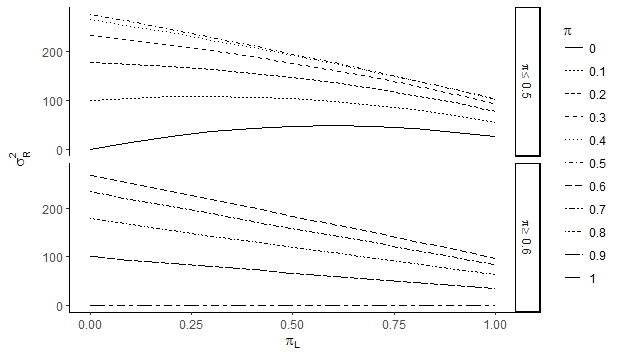} 
\caption{Cross-sections of $E(\varr)=g(\pi, \pi_L)$ for fixed values of $\pi$}\label{graphVarRCrossSections}
\end{figure}

A graph of cross-sections of $E(\varr)=g(\pi, \pi_L)$ in equation (\ref{ratiovarline}) for $\pi =0, 0.1, 0.2,...,1.0$ is shown in Figure \ref{graphVarRCrossSections}.  The only real-valued critical point of $g$ requires $\pi_L=2.41 \notin [0,1]$.  Thus, $g$ will be maximized on its boundary.  The maximum value of $g$ on its boundaries is shown in Table \ref{boundaryratio}. Thus, we find the max of $g(\pi,\pi_L)=E(\varr)=261$ occurs at $\pi_L=0$ and $\pi=0.5$

\begin{table}
\centering
\begin{tabular}{cc}
\hline 
Boundary & Max of $g(\pi,\pi_L)=E(\varr)$\\ 
\hline 
$\pi=0$ & 44 \\ 
$\pi=1$ & 0 \\ 
$\pi_L=0$ & 261 \\ 
$\pi_L=1$ & 102 
\end{tabular}
\caption{Maximum values of $E(\varr)$ on Domain Boundaries}
 \label{boundaryratio}
\end{table}

\section{Conclusions and Further Research}
Using a binomial generative model and assuming all-or-nothing errors, we developed a method for choosing between ratio and simple expansion estimators. We  showed that, for any audit population, ratio estimation is likely to outperform simple expansion, provided the assumptions for ratio-estimator-based confidence intervals are valid. We further extended existing estimates of $\varyy$ and $\varr$ under the all-or-nothing error assumption to a novel, realistic partial error model based on line-item errors. Notably, the methods we have developed can be implemented without pilot study data, requiring only the known claim data and estimated error rate(s).  Moreover, in the absence of estimated error rate(s), conservative sample sizes can be calculated by maximizing the variance over the error rate, $\pi$, for all-or-nothing errors populations or over $(\pi,\pi_L)$ for partial error populations.

We have also demonstrated that optimal stratification under ratio estimation is unaffected by $\pi$ so long as $\pi$ can be assumed to be uniform across the claim population. Although ratio estimation has been shown to have many desirable properties, it is also known that ratio-estimator-based confidence intervals may fall short of the nominal confidence level in audit populations. Thus, it would be useful to investigate whether the estimator of $\varr$ in equation(\ref{robertsratio}) or equation (\ref{ratiovarline}) would improve the attained confidence level over the standard estimate, $\hat{\sigma}_R^2 = \sum_{i=1}^n (Y_i-\hat{R}X_i)^2/(n-1)$. Finally, since the generative models in this paper all assume that the probability of a claim being in error is independent of the claim amount, it would be useful to extend these results to a generative model which allows for dependency between the probability of an error and the claim amount.\\

\bigskip

\bibliographystyle{asa}

\bibliography{auditrefs}

\end{document}